\documentclass[10pt ]{IEEEtran}

\usepackage{graphicx}
\usepackage{subcaption}

\makeatletter
\def\@makefnmark{%
  \leavevmode
  \raise.9ex\hbox{\fontsize\sf@size\z@\normalfont\tiny\@thefnmark}}
\makeatother

\begin{document}
\pagestyle{empty}

\title{A Scalable Mobility-Centric Architecture for Named Data Networking}

\author{\IEEEauthorblockN{ Aytac Azgin, Ravishankar Ravindran, Guoqiang Wang}\\ 
\IEEEauthorblockA{Huawei Research Center, Santa Clara, CA, USA}\\
\{aytac.azgin,ravi.ravindran,gq.wang\}{@huawei.com}}
\maketitle
\thispagestyle{empty}

\begin{abstract}
Information-centric networking (ICN) proposes to redesign the Internet by replacing its host centric design with an information centric one, by establishing communication at the naming level, with the receiver side acting as the driving force behind content delivery. Such design promises great advantages for the delivery of content to and from mobile hosts. This, however, is at the expense of increased networking overhead, specifically in the case of Named-data Networking (NDN) due to use of flooding for path recovery. In this paper, we propose a mobility centric solution to address the overhead and scalability problems in NDN by introducing a novel forwarding architecture that leverages decentralized server-assisted routing over flooding based strategies. We present an in-depth study of the proposed architecture and provide demonstrative results on its throughput and overhead performance at different levels of mobility proving its scalability and effectiveness, when compared to the current NDN based forwarding strategies.
\end{abstract}


\IEEEpeerreviewmaketitle

\section{Introduction}

\emph{Information-centric Networking} (ICN) is a new networking paradigm that addresses the shortcomings of the current Internet architecture by shifting the focus from the host-centric communication model to a content-centric one \cite{SurveyICN}. ICN uses a unique --flat or hierarchical-- naming convention to name content, which represents the main driving force for information dissemination.

Multiple architectures have so far been proposed to guide the development of ICN (see \cite{ICNsurvey12,MainICNSurvey} for a detailed overview). Architectures for information-centric networks are uniquely defined by how they handle \emph{naming} and \emph{name resolution}. Our research focuses on one of those proposals, namely the \emph{Named-data Networking} (NDN) proposal, and addresses one of its major concerns, namely the \emph{mobility}.

NDN assumes hierarchically structured names (consisting of any number and size of \emph{components}) to support scalable routing and utilizes request/response type message (referred to as \emph{Interest} and \emph{Data}) processing at every hop. In NDN, each packet carries a name that can be used to identify a content, service, host, or user. Content authenticity is provided by using digital signatures that are delivered with the content. Due to its flexibility to use broadcast medium efficiently and to support loop-free forwarding on multiple interfaces, NDN is a good match to deliver (or acquire) content to (or from) mobile hosts.

However, despite the inherent advantages NDN possesses to support ad hoc networking by resolving content requests to location using online forwarding strategies, overhead associated with re-routing Interests to mobile hosts can be overwhelming, thereby, limiting NDN's efficiency for especially delivering mobile-originated content. Thus, mobility can be considered as a major obstacle in creating a scalable network architecture based on NDN.

Mobility concerns for information-centric networks have generally been addressed by introducing location-aware routing to content delivery. However, these solutions typically assume the presence of \emph{Global Resolution Servers} to handle the host mobility or work on the flat-name space (\emph{e.g.}, \cite{DONA,NetInf}), hence, they are not directly applicable to NDN, which fundamentally does not distinguish between an entity identifier and its location. In \cite{385} the authors briefly suggest the use of \emph{forwarding hints} as a guideline to enable location-driven forwarding in named-data networks, which represents our starting point to develop a comprehensive mobility solution to supplement the current NDN architecture.

We can state our contributions in this paper as follows. We propose a scalable and stable mobility solution for the NDN architecture based on location-centric forwarding, by separating content names and network addresses to enable \emph{iterative-binding}. Proposed solution uses a decentralized resolution architecture, \emph{i.e.}, distributed \emph{name resolution servers}, along with in-band mobility awareness of named entity to provide dynamic name-location mappings, which in turn enables quick-recovery of the named-data path during mobile-driven content delivery. The proposed solution achieves resource efficient forwarding by avoiding \emph{network flooding} during content delivery after handovers. The proposed solution is backward compatible as it allows the proposed extensions to be applied to mobile entities, while non-mobile entity traffic can be treated in the standard manner; this allows mobility to be treated as a \emph{service}. The extensions also support policy-based routing by controlling intra- and inter-domain routing within the network, in the sense that the proposed data structures can be invoked only if required compared to today's situation where all mobile entities are treated alike.

To evaluate our solution, we developed extensions to the \emph{ndnSIM} simulator \cite{ndnSIM} 
and analyzed the performance of the proposed solution by comparing it to the default mobility-driven NDN forwarding policies (\emph{i.e.}, \emph{flooding-based policies}). We tested the proposed solution under various network topologies and mobility scenarios and observed significant improvements in the scalability performance while achieving comparable performance, in throughput, to the current NDN forwarding policies.

The rest of the paper is organized as follows. In Section~\ref{Section:NDN} we briefly explain the NDN architecture and mobility solution for the NDN. We present the proposed forwarding architecture in Section~\ref{Section:Approach}. We analyze the performance of our solution in Section~\ref{Section:Analysis}. We discuss the practical considerations addressing scalability, storage, and security concerns in Section~\ref{Section:Discussions}. Section~\ref{Section:Conclusion} concludes our paper.

\section{Mobility in Named-data Networks}\label{Section:NDN}

In NDN, to support name-based routing, each node is equipped with three components: ($i$) \emph{Content Store} (CS), ($ii$) \emph{Pending Interest Table} (PIT), and ($iii$) \emph{Forwarding Information Base} (FIB). CS is the local cache used to store content in an NDN router. Anytime a host receives an Interest for a locally cached content, a Data packet is created in response and forwarded through the incoming interface(s) for the Interest\footnote{NDN uses the term \emph{face} to represent the interface over which a packet is received or delivered.} before the Interest is discarded. PIT stores the set of active (or pending) Interests forwarded by the host and waiting for the corresponding Data to be delivered. PIT entries track the incoming faces for the received Interests and the entries are created per content, \emph{i.e.}, any subsequent Interest for an active PIT entry is suppressed at the local host, and the current PIT entry is updated with the incoming face information. 
Also note that, PIT helps prevent the formation of routing loops. For that purpose, each Interest carries a random nonce to detect duplicate requests.
Anytime a host receives a matching Data for a pending Interest, received Data packet is forwarded along the incoming face(s) indicated by the PIT entry, before the request is removed from the PIT. FIB aggregates forwarding information at each host, and consists of entries mapping content names to outgoing faces. To select the outgoing faces matching the prefix of an Interest, FIB uses the longest prefix matching.

To handle mobility, NDN typically relies on overhead-heavy flooding based strategies, which introduces major scalability concerns \cite{AzginICC14}. The authors in \cite{385} address this problem by proposing the use of forwarding hints to direct requests to the content source. For that purpose, the authors aim to bring the hierarchical DNS structure to NDN, by proposing separate name and resolution servers to provide the necessary mappings (see \cite{AlexThesis} for details). However, because of the iterative/recursive lookups, the proposed approach may introduce non-negligible latency (multiple RTTs), and synchronization may be needed to support \emph{entity} (\emph{i.e.}, user, device, content, or service) mobility.

Our solution, on the other hand, expands the core NDN architecture and creates a mobility-based solution from within, by relying on a completely decentralized architecture, where all the operation follows the NDN principles and utilize only Interest/Data exchanges to support host mobility. The principles of our solution are listed as follows:
\begin{itemize}
\item \emph{Forwarding scalability}: FIB size should be independent of the number of mobile entities. We can achieve this by requiring to keep object state only at the edges (\emph{i.e.}, gateway points), and perform routing at the core network based on locator prefixes.

\item \emph{Control overhead}: Updates due to mobility should be local and should not affect routing/forwarding convergence. We can achieve this by using a \emph{Local Controller} (within each domain/AS) that is capable of resolving any mobile entity using its home binding.

\item \emph{Intra-} and \emph{Inter-session mobility}: The architecture should support both intra- and inter-session mobility (\emph{i.e.}, changing location during a session or between sessions, where the term \emph{session} refers to the delivery of a whole content, file, video, etc.). While \emph{inter-session mobility} is handled through an update to/from the \emph{Home Controller}, \emph{intra-session mobility} is enabled by introducing a \emph{Mobility-Update} tag, which triggers an update at all the related consumer-attached Service Routers to re-resolve the locator for the mobile entity.

\item \emph{Mobility granularity}: As producer mobility incurs cost in terms of control infrastructure, it should be realizable as a \emph{service}, hence granularity to support mobility of any \emph{entity} should be supported. Further differentiation can be supported in terms of geographic span of mobility, latency, etc.
\end{itemize}

In short, our solution is complimentary to NDN to handle mobility in a scalable manner. The enhancements are optional, hence, should be compatible with existing implementations as well. Next, we present our architecture.

\section{Proposed Architecture}\label{Section:Approach}

\begin{figure}[htb]
  \centering
  \includegraphics[height=2.5in, width=1.5in,angle=270]{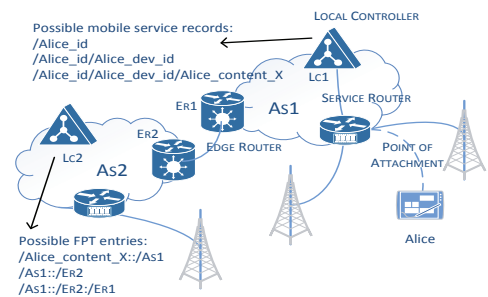}\\
  \caption{ICN mobility example.}\label{figure:initial_example}
\end{figure}

\begin{figure}[htb]
  \centering
  \includegraphics[height=2.5in,angle=270]{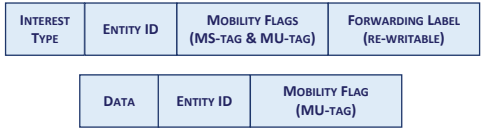}\\
  \caption{Updated protocol data unit formats with \emph{mobility flag}s and/or \emph{forwarding label}.}\label{figure:pdu}
\end{figure}

Leveraging ICN's name-based mobility, the proposed architecture allows any meaningful space of the name hierarchy to be mobile. For instance, in Figure~\ref{figure:initial_example}, Alice's name space can be recorded as follows. Alice can choose all of her devices, ``/Alice\_id/", or one of her devices, ``/Alice\_id/Alice\_dev\_id", to be mobile, or a subset of her device's content space, ``/Alice\_id/Alice\_dev\_id/Alice\_content\_X", to be mobile. User enables this by actively registering this name space to the network, where the mobility service enabled by the provider allows the entities under the name space to be accessible anywhere on the Internet. This motivates a scalable mobility architecture. The proposed solution achieves this objective by utilizing four essential building blocks: \emph{Local Controller}s to provide name-to-locator mappings and manage control overhead, \emph{Fast Path Table}s (which are used by the designated routers to control information flow in the network) to address forwarding scalability, \emph{Forwarding Label}s and \emph{Mobility Tag}s to address inter- and intra-session mobility at different mobility granularities. We observe the resulting Interest/Data packet formats in simplified view in Figure~\ref{figure:pdu}, which shows the additional components integrated into each packet format.

We consider a basic networking hierarchy, which consists of a given number of domains or \emph{Autonomous Systems} (\textsc{As}). Each domain/\textsc{As} is assigned a \emph{Local Controller} (\textsc{Lc}) service carrying a domain designated prefix, \emph{e.g.}, \textsc{Lc} for \textsc{As}-(ID:11) is assigned the prefix ``\textsc{/LocalController:As11}". \textsc{Lc}s are responsible for mapping local and remote entity names to domain/router identifiers. If the entity is local (\emph{intra-}Domain or \emph{intra-\textsc{As}} delivery), it resolves to \emph{Service Router} (which can also act as a \emph{Point of Attachment}\footnote{In the paper, we separate the \emph{Point of Attachment} and the \emph{Service Router} to clearly identify the different functionalities assigned to each by the proposed framework.}), whereas, if the entity is remote (\emph{inter-}Domain or \emph{inter-\textsc{As}} delivery), it maps to local domain's egress router identifier. 
Each host is  statically or dynamically assigned to a designated \textsc{Lc} (referred as \emph{Home Controller}), which stores up-to-date information regarding its users (we will explain shortly the process to acquire such information). Also note that, proposed system allows-and supports for-the endpoints to change their Home Network bindings on-the-fly, during content delivery.

Per the NDN requirements, hierarchical names are used to identify the entities\footnote{Hereafter, we will use entity/content interchangeably.}. In the following, we pursue our discussion with respect to host mobility, which can be generalized to mobility of other named entities too. Each host is assigned a unique identifier representing the association of a user to its home network (\emph{e.g.}, ``\textsc{/As:Home/Dev:Id}" with ``Dev:Id" also including the ``Host:Id"). Complete identifier, \emph{if available}, is only needed during \textsc{Registration}. Network identifier, for an endpoint, is not considered \emph{a priori} requirement to support successful location discovery. Using an identifier, however, is the preferred choice to minimize control overhead.

We assume an \textsc{Lc} to have access to information directly related to communication taking place within its domain (\emph{e.g.}, home controller or locator information on content being published within the domain). Hence, no direct \emph{synchronization} is assumed to exist among the \textsc{Lc}s. Active domain information on visiting hosts is flushed on a regular basis, as soon as the host leaves the controller's domain, to minimize access to outdated information. However, we allow the \emph{Remote Controller} to store information on \emph{Home Controllers} representing the visiting hosts for longer periods to minimize overhead associated with the \emph{Initial Discovery Phase}.

We utilize \emph{Forwarding Labels} to route packets in a controlled manner. Note that, routing is only needed to deliver the Interest packets, as Data packets follow the reverse path, unless tunneling is used to alter the Data path. Forwarding label is similar to content prefix in the sense that, when used, it provides sufficient information on the next physical or logical hop (\emph{i.e.}, gateway points). However, unlike a content prefix, forwarding label is used as a dynamic tag within the Interest that is regularly updated (at the service routers and gateway points) along the path to content source. To support the use of forwarding labels, at each service or edge router, we utilize a \emph{Fast Path Table} (FPT) that carries the mappings corresponding to content prefixes and forwarding addresses.

Since handling mobility with FPT and forwarding labels incurs memory and computational cost, forwarding decisions based on FPT can be limited to traffic tagged as being part of the \emph{Mobility Service}, whereas for the other services, default routing based on FIB can be used. Specifically, because of the overhead associated with the use of forwarding labels, we consider its implementation as a \emph{Service} (which we call as the \emph{Forwarding Label-Service}, or the \emph{FL-Service}) and limit its use to mobile-only scenarios. Furthermore, any mobile user that wishes to use the \emph{FL-Service} indicates its intent during \emph{\textsc{Registration}} by setting a single-bit \emph{Mobility Service} tag (\emph{MS-tag}) contained within the registration message. In addition, a consumer can also enable this tag along with the mobile entity's unique name to invoke the \emph{FL-Service}, which helps an NDN router differentiate between mobile and non-mobile entities. In scenarios where the \emph{FL-Service} is not explicitly defined, forwarding is strictly based on the core NDN policies. \emph{FL-Service}, however, is capable of taking full advantage of NDN's strengths, \emph{i.e.}, in-network caching and the available FIB entries.

\begin{figure}[htb]
  \centering
  \includegraphics[height=3in, width=1.8in,angle=270]{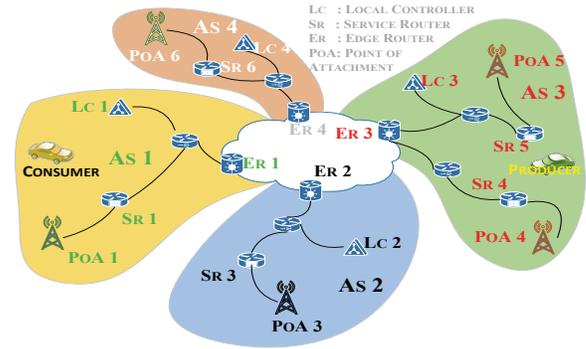}\\
  \caption{A simple networking scenario with two mobile endpoints (one \emph{Consumer} and one \emph{Producer}) and four Autonomous Systems (ASs).}\label{figure:networking_scenario_1}
\end{figure}

The proposed framework and the steps taken to initiate content delivery between the involved parties are explained using the scenario shown in Figure~\ref{figure:networking_scenario_1}. For the given example, we use four Autonomous Systems with a single Consumer node, residing in \emph{\textsc{As1}} and a single \emph{Producer} node, residing in \emph{\textsc{As3}}. We assume that \emph{\textsc{As2}} represents the Home Network for the \emph{Producer} node. We also assume that \emph{Producer} has already registered with its Home Network and was assigned the identifier ``\textsc{/As2/Producer:Id/}". Finally, we assume that \emph{Producer} initiates the remote registration phase during its attachment process in the foreign domain as in \emph{\textsc{As3}}. Also, note that, we explain our solution, from the perspective of a mobile \emph{Producer}, which is the more challenging scenario. Additional modifications needed for the mobile \emph{Consumer} case are omitted as they share similar traits to the mobile \emph{Producer} case, in regards to local updates, to ensure that stale information at the \emph{Service Routers} and \emph{Local Controllers} are promptly taken care of.

\subsection{Registration Phase}

\begin{itemize}
\item \emph{\textsc{Step i}}: \emph{Registration Phase} initiates with \emph{Producer} sending a \textsc{Register} message for its content, referred simply as \textsc{/Prefix}, to its host network's \textsc{Lc} by also including its complete identifier ``\textsc{/Prefix:/As2/Producer:Id}", where \textsc{/As2/Producer:Id} represents the home binding identifier for the entity \textsc{/Prefix}. Here, \textsc{Producer:Id} can be the device-ID, and if the device itself is mobile then \textsc{/Producer:Id} can be considered as part of \textsc{/Prefix} (hence, ``\textsc{/As2}" is good enough for the home binding). Furthermore, the \textsc{/Prefix} itself can be globally routable or not, as it is resolved using the \textsc{Lc} infrastructure. In short, assuming that the hosts learn the minimum control name space to interact with the \textsc{PoA}s during mobile attachment, \emph{Producer} sends a registration message to \textsc{PoA5} with the prefix ``\textsc{/PoA5/Reg}"\footnote{For the sake of simplicity, we added \textsc{As} identifier to the control prefix. However, a single \textsc{Lc} prefix, common to all domains, can be used to route Interests to a single \textsc{Lc} within each domain, as these Interests will not be forwarded outside the domain.}\textsuperscript{,}\footnote{Depending on the implementation, information on published content is either included within the registration prefix as ``\textsc{/PoA5/Reg:$<$/Prefix$>$}", or separately within the forwarding label.}. Using the FIB/FPT entries, registration message is forwarded towards \emph{\textsc{Lc3}} through \emph{\textsc{PoA5}} and \emph{\textsc{Sr5}}. \emph{\textsc{Sr5}} also updates the forwarding label, by including its address to inform \emph{\textsc{Lc3}} on the identity of the Service Router associated with the \emph{Producer}. After \emph{\textsc{Lc3}} receives the registration message, if the \emph{MS-tag} is set, local database (\textsc{L-db}) is updated with the entry ``\textsc{/Prefix::/Sr5:Address::/Home:As2}"\footnote{``\textsc{:Adress}" can refer to physical address space or name space. Hereafter, we drop the ``\textsc{:Adress}" postfix from the host name and assume router name to represent any such address.}, where the third component specifies the Home Network for ``\textsc{/Prefix}". Note that, FPT or \textsc{L-db} entries by default require at least two inputs, \emph{prefix} information and \emph{locator} information. We use double colon to separate the entries. Similarly, if the \emph{MS-tag} is set in the registration message, FPT tables at \emph{\textsc{PoA5}} and \emph{\textsc{Sr5}} are also updated, \emph{e.g.}, by adding the entry ``\textsc{/Prefix::/Producer}" to FPT table at \emph{\textsc{PoA5}}, and the entry ``\textsc{/Prefix::/PoA5}" to FPT table at \emph{\textsc{Sr5}}, upon receiving the acknowledgement from \textsc{Lc3}.

\item \emph{\textsc{Step ii}}: \emph{Proactive Update Phase} initiates with \emph{\textsc{Lc3}} sending a \emph{\textsc{Route-Update}} message to the Edge Routers (\textsc{Er}s) (\emph{\textsc{Er3}}, for the given scenario) to update their FPT table with the entry ``\textsc{/Prefix::/Sr5}" so that any Interest received by the \textsc{Er}s targeting ``\textsc{/Prefix}" can be immediately forwarded to \emph{\textsc{Sr5}}. Note that, at the \textsc{Er}s, in addition to proactive updates, we can also use reactive updates, with the \textsc{Er} making a request for the forwarding information. However, it requires the initial traffic to be forwarded to the \textsc{Lc}s to minimize latency, and additional overhead to keep track of entries to avoid sending recurring route requests for non-local \emph{Producers}.

\item \emph{\textsc{Step iii}}: \emph{Foreign Home Registration Phase} initiates with \emph{\textsc{Lc3}} sending \emph{\textsc{Lc2}} a \textsc{Home-Register} (\textsc{Hreg}) message for ``\textsc{/Prefix}", once again using the complete identifier info for \emph{Producer}. Prefix for the \textsc{Hreg} message is given by ``\textsc{/LocalController:As2/Hreg}". After \emph{\textsc{Lc2}} receives the registration message, it updates its \textsc{L-db} as follows:
    \begin{itemize}
    \item If an active entry is found in \textsc{Lc2}'s \textsc{L-db} for \emph{Producer}, corresponding to a different \textsc{As}, say \emph{\textsc{As4}}, the entry is updated as follows: ``\textsc{/Prefix::/Remote:As3::/Remote:As4}", where the third component indicates the previous \textsc{As} the \emph{Producer} was in. Additionally, \emph{\textsc{Lc2}} sends a \textsc{Flush-Register} message to \textsc{Lc4} (which will be explained shortly).

     \item If no active entry is found in \emph{\textsc{Lc2}}'s \textsc{L-db} for \emph{Producer} or \emph{Producer} was previously in \emph{\textsc{As2}}, \textsc{L-db} at \textsc{Lc2} is updated with the following entry: ``\textsc{/Prefix::/Remote:As3::/}".
    \end{itemize}
\end{itemize}

Note that, all communication in the network utilize the core Interest-Data message exchange procedures as defined by NDN. Hence, to support a reliable control message exchange, each registration/update message is followed by an acknowledgement. By jointly utilizing (content) name and (forwarding) label fields in the Interest, we can deliver all the necessary information to \textsc{Lc}s.

\subsection{Content Delivery Phase}

Assume that \emph{Consumer}, who is currently connected to \emph{\textsc{PoA1}}, wants to receive ``\textsc{/Prefix}" published by {\emph{Producer}}. Also assume that no \emph{\textsc{Sr}} or \emph{\textsc{Lc}} in \emph{\textsc{As1}} has any FIB/FPT/\textsc{L-db} entry for ``\textsc{/Prefix}".

\begin{itemize}
\item \emph{\textsc{Step i}}: \emph{Consumer} starts \emph{Content Delivery Phase} by sending an Interest for ``\textsc{/Prefix}" to \emph{\textsc{PoA1}}, who then checks, in order, its CS and PIT to find a matching entry for ``\textsc{/Prefix}. Since no entry is found, \emph{\textsc{PoA1}} forwards the Interest to \emph{\textsc{Sr1}}. After receiving the Interest, \emph{\textsc{Sr1}} performs a more detailed check including searching for FPT (and, if necessary FIB) entries. Per our assumption, since no entry is found, \textsc{{SR1}} creates a {\textsc{Route-Request}} (\textsc{Rreq}) message with the prefix ``\textsc{/LocalController:As1/Rreq}" and sends it to \emph{\textsc{Lc1}}. Furthermore, \emph{\textsc{Sr1}} updates its local \emph{{request waiting list}} (\textsc{Rwl}) for the request messages it creates (similar to the PIT entries) to avoid retransmitting the same request to \textsc{Lc1}. Any further Interests targeting ``\textsc{/Prefix}" are queued at \emph{\textsc{Sr1}} until a response to the first \textsc{Rreq} message is received.

\item \emph{\textsc{Step ii}}: After \emph{\textsc{Lc1}} receives the \textsc{Rreq} message, it searches for a matching entry within its \textsc{L-db}. If an entry is found, request can be \emph{unicast} to the \textsc{Lc} of the home domain. If no entry is found, and the received content name does not identify the home binding (\emph{i.e.}, domain-based association), in a format similar to ``\textsc{/Prefix:/As2/Producer:Id}", then the request is forwarded to all the other \textsc{Lc}s. For the given scenario, \emph{\textsc{Lc1}} forwards \textsc{Rreq} to \emph{\textsc{Lc2}}, \emph{\textsc{Lc3}} and \emph{\textsc{Lc4}}. We can reduce the discovery overhead by using a controlled name-based multicast, \emph{i.e.}, by grouping multiple requests into a single Interest and forwarding the Interest \textsc{As}-by-\textsc{As} until a match is found. However, in practice, we can expect such home mapping to be provided at the time of request. Also, similar to how it was with \emph{\textsc{Sr1}}, \emph{\textsc{Lc1}} also implements a local \textsc{Rwl} for the received \textsc{Rreq} messages.

\item \emph{\textsc{Step iii}}: After \emph{\textsc{Lc2}} receives the \textsc{Rreq} message, a Data packet is created with the mapping ``\textsc{/Prefix::/As3}" that points to the current location of the \emph{Producer}.

\item \emph{\textsc{Step iv}}: After \emph{\textsc{Lc1}} receives \emph{\textsc{Lc2}}'s response to its \textsc{Rreq} message, \textsc{Lc1} first updates its \textsc{L-db} with the entry ``\textsc{/Prefix::/As3::/Home:As2}". \emph{\textsc{Lc1}} also creates a Data packet in response to the received \textsc{Rreq} messages and alerts any \textsc{Sr} indicated by the \textsc{Rwl} entry associated with ``\textsc{/Prefix}", before removing such entries. For the given scenario, \emph{\textsc{Lc1}} responds to \emph{\textsc{Sr1}}'s \textsc{Rreq} message with the mapping ``\textsc{/Prefix::/As3::/Er1}", by also including information on the egress point, \emph{\textsc{Sr1}} needs to use to reach \emph{\textsc{As3}} from \emph{\textsc{As1}}\footnote{Depending on the inter-domain routing policies, \textsc{Lc1} can also provide a complete path information for the Interests to go from \textsc{As1} to \textsc{As3}, minimizing the processing overhead at the egress points.}.

\item \emph{\textsc{Step v}}: After \emph{\textsc{Sr1}} receives a response to its \textsc{Rreq} message, \textsc{Sr1} updates its FPT with the following entries:  ``\textsc{/Prefix::/Remote:As3}" and ``\textsc{/As3::/Er1}". \emph{\textsc{Sr1}} then clears its \textsc{Rwl} entry for ``\textsc{/Prefix}" and starts forwarding the corresponding Interests using the forwarding label ``\textsc{/Er1/Remote:As3}".

\item \emph{\textsc{Step v}}: After \emph{\textsc{Er1}} receives an Interest, it checks the forwarding label and determines \emph{\textsc{As3}} as the targeted domain. We assume that the \textsc{Er}s have already populated their FPTs with mappings of ``\textsc{/Remote:As::/NextHop-Er}", thereby allowing \emph{\textsc{Er1}} to set the received Interest's forwarding label to ``\textsc{/NextHop-Er/Remote:As3}". This process is repeated until the Interest is received by the gateway point to \textsc{As3}, that is, \emph{\textsc{Er3}}.

\item \emph{\textsc{Step vi}}: After \emph{\textsc{Er3}} receives the Interest, it determines that the target for the received Interest resides within its domain (\emph{i.e.}, \emph{\textsc{As3}}). \emph{\textsc{Er3}} uses its FPT to determine the address for the \textsc{Sr} servicing the \textsc{PoA} connected to \emph{Producer} with ``\textsc{/Prefix}". FPT-based lookup process is repeated at \emph{\textsc{Sr5}} (and \emph{\textsc{PoA5}}, \emph{if necessary}) until \emph{Producer} receives the Interest, after which content delivery towards \emph{Consumer} can proceed along the reverse path, using NDN's breadcrumb approach.
\end{itemize}

\subsection{Intra-AS Handover Phase}

Now, assume that \emph{Producer} moves from \emph{\textsc{PoA5}}'s service area to \emph{\textsc{PoA4}}'s service area, triggering an intra-\textsc{As} handover.

\begin{itemize}
\item \emph{\textsc{Step i}}: After the handover to \emph{\textsc{PoA4}} completes, \emph{Producer} sends a \emph{\textsc{Register}} message to \emph{\textsc{Lc3}}, as explained earlier. After \emph{\textsc{Lc3}} receives the \emph{\textsc{Register}} message, it looks up for a matching entry for ``\textsc{/Prefix}" in its \textsc{L-db} and notices that \emph{Producer} is previously associated with a different \textsc{Sr}. \emph{\textsc{Lc3}} updates its \textsc{L-db} entry with ``\textsc{/Prefix::/Sr4::/Home:As2}" and sends ($i$) a \emph{\textsc{Route-Update}} (\textsc{Rupd}) message to the \textsc{Er}s (\emph{i.e.}, \emph{\textsc{Er3}} for the given scenario) to update their FPT tables with the entry ``\textsc{/Prefix::/Sr4}", and ($ii$) a \emph{\textsc{Flush-Register}} (\textsc{Freg}) message to \emph{\textsc{Sr5}}.

\item \emph{\textsc{Step ii}}: After \emph{\textsc{Er3}} receives the \emph{\textsc{Rupd}} message, it updates its FPT table and forwards any matching Interest towards \emph{\textsc{Sr4}}.

\item \emph{\textsc{Step iii}}: After \emph{\textsc{Sr5}} receives the \emph{\textsc{Freg}} message for ``\textsc{/Prefix}", it updates the local FPT with the entry ``\textsc{/Prefix::/Sr4}" and forwards any new Interest targeting \emph{Producer} to \emph{\textsc{Sr4}} (note that, \textsc{Sr5} can also forward any undelivered Interest to \textsc{Sr4}). \emph{\textsc{Sr5}} also starts a timer to flush its local entry. Additionally, \emph{\textsc{Sr5}} forwards the \emph{\textsc{Freg}} message to \emph{\textsc{PoA5}}, which flushes its FPT entry upon receiving.
\end{itemize}

Even though the default process for \emph{de-registration} is to go through \textsc{Lc}s, our framework also allows for \emph{de-registration} through the \textsc{PoA}s, to further improve the latency performance.

\subsection{Inter-AS Handover Phase}
For the inter-\textsc{As} handover, now assume that \emph{Producer} moves from \emph{\textsc{PoA5}}'s servicing area to \emph{\textsc{PoA6}}'s servicing area, hence leaving \emph{\textsc{As3}} to join \emph{\textsc{As4}}.
\begin{itemize}
\item \emph{\textsc{Step i}}: After \emph{Producer} moves to \emph{\textsc{As4}}, it initiates the registration phase by sending the \emph{\textsc{Register}} message upstream towards \emph{\textsc{Lc4}}, and triggering updates along the path to (and at) \emph{\textsc{Lc4}} (at \emph{\textsc{PoA6}} and \emph{\textsc{Sr6}}).

\item \emph{\textsc{Step ii}}: \emph{\textsc{Lc4}} sends a \emph{\textsc{Home-Register}} message to \emph{\textsc{Lc2}}, while also sending a \emph{\textsc{Route-Update}} message to \emph{\textsc{Er4}}, which updates its FPT with the entry ``\textsc{/Prefix::/Sr6}".

\item \emph{\textsc{Step iii}}: After \emph{\textsc{Lc2}} receives the \emph{\textsc{Hreg}} message, it determines a change-of-domain for the \emph{Producer}, moving from \emph{\textsc{As3}} to \emph{\textsc{As4}}, and sends \emph{\textsc{Lc3}} a \emph{\textsc{Freg}} message with the prefix ``\textsc{/LocalController:As3/Freg}". \emph{\textsc{Lc2}} also includes information on the currently active domain (\emph{i.e.}, \emph{\textsc{As4}}) for \emph{Producer} in its message to \emph{\textsc{Lc3}}.

\item \emph{\textsc{Step iv}}: After \emph{\textsc{Lc3}} receives the \emph{\textsc{Freg}} message, it creates a local \emph{\textsc{Freg}} message that it sends to \emph{\textsc{Sr5}}. {\textsc{Lc3}} next sends a \textsc{Route-Update-With-Timeout} message to \textsc{Er3} requiring it to update its FPT entry to point to the new domain for ``\textsc{/Prefix}" for a forwarding-timeout period (depending on an estimate for the recovery timeframe). Anytime \textsc{Er3} receives an Interest targeting ``\textsc{/Prefix}" with the wrong forwarding label (\emph{i.e.}, pointing to \textsc{As3}) during the FPT-timeout interval, the timeout parameter is reset to its default value (to ensure any \emph{Consumer} not aware of domain change is informed).
    \begin{itemize}
    \item The forwarding label for the incorrectly labeled Interest is updated with the correct label, and the Interest's \textsc{Mobility-Update} tag is set to 1, before the Interest is forwarded towards the correct domain by the \textsc{Er} (\emph{i.e.}, \textsc{As4} for the given example). In the case of a failure related to previous network's \textsc{Er}, hard timeouts are used at the \emph{Consumer} side to force a location update through the \textsc{Lc}.
    \item When \emph{Producer} receives an Interest with the \emph{MU-tag} set, suggesting that the \emph{Consumer}'s network is not aware of \emph{Producer}'s change-of-domain, it sets the \emph{MU-tag} within the Data packet as well.
    \item When \textsc{Sr1} (or any relevant consumer serving \textsc{Sr}) receives a Data packet with the \emph{MU-tag} set, \textsc{Sr1} initiates the \emph{Re-Discovery Phase} by requesting a forced \textsc{Rupd} from \textsc{Lc1}, which then contacts \textsc{Lc2} to acquire the up-to-date domain information. Another alternative to the above approach is for the \emph{Producer} side to include domain information within Data packets, in response to an Interest with a set \emph{MU-tag}.
    \end{itemize}
\end{itemize}

\section{Performance Analysis}\label{Section:Analysis}
\begin{figure}[htb]
  \centering
  \includegraphics[height=2.2in, width=1.7in,angle=270]{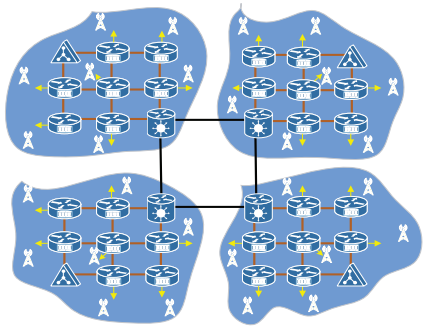}\\
  \caption{``2x2 AS with 3x3 intra-AS" Grid network topology with 4AS.}\label{figure:networking_scenario_2}
\end{figure}

We implemented the proposed architecture in ndnSIM \cite{ndnSIM}, by designing the modules and applications required for the processing of the received packets at designated hosts. We approximated the mobile handover by enabling/disabling wireless network interfaces based on the respective signal strength for the channel between a host and the access points reachable by the host. We assumed a handover latency of $50ms$\footnote{Note that, handover latency depends on many factors, including the wireless technology utilized by the host and the type of handover initiated by the mobile host, resulting in latencies of tens-to-hundreds of ms. The chosen close-to-optimal value for the handover latency allows us to specifically focus our attention on the impact of location change.}. For the point-to-point (wireline) links, we assumed a bandwidth of $10Mbps$ and a propagation latency of $10ms$. We used constant-bit-rate (CBR) traffic model with a request rate of $20$ packets per second. In our simulations, we considered grid-based topologies of varying sizes (\emph{i.e.}, 4AS scenario with $64$ nodes, 9AS scenario with $137$ nodes, and 16AS scenario with $236$ nodes)\footnote{Grid-based topology is chosen due to its flexibility in proportionally extending the network size and measuring its impact on perceived performance.}. We show the 4AS scenario in Figure~\ref{figure:networking_scenario_2}. The 2nd and the 3rd scenarios, not illustrated here for brevity purposes, are an extension of the 1st scenario, with 3x3 and 4x4 AS formations.

We compared the performance of our forwarding solution to the \emph{Flooding} and \emph{Semi-flooding} (which is a modified version of \emph{Smart-flooding} that uses Flooding in the access network to minimize the impact of timeouts) techniques (our approach utilizes the \emph{Best-route} strategy whenever needed). In a network with unlimited resources, Flooding technique, which continuously flood the network, achieves the best performance in the end-to-end throughput, while achieving the worst overhead performance. Semi-flooding, on the other hand, uses intelligent flooding only after losses, hence represent the tradeoff between resource efficiency and end-to-end throughput.

\begin{figure}[htb]
  \centering
  \includegraphics[height=3.3in, angle=270]{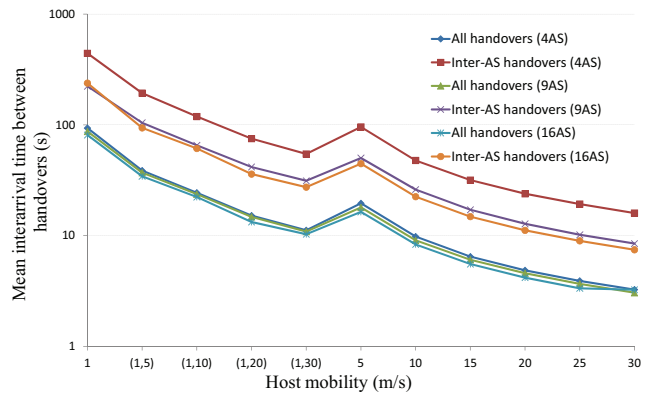}\\
  \caption{Interarrival times for the mobile handovers for the three considered network topologies, also including the results for the inter-AS handovers ($\approx20$--$34\%$ of the handovers are inter-AS handovers).}\label{figure:handover_results}
\end{figure}

Due to space limitation, we share our simulation results\footnote{We ran $10$ simulations using different random seeds, each of which lasts for $30$ minutes, and present the average of their results.} for a single \emph{Consumer}-\emph{Producer} pair, which is sufficient to illustrate the differences between the three approaches, and specifically focused on the \emph{Producer} mobility (\emph{i.e.}, \emph{Consumer} is assumed to move at the lowest mobility level, whereas the \emph{Producer} mobility is varied from low-to-high\footnote{\emph{Consumer}'s mobility region is limited to the top row of ASs, whereas the remaining ASs are used to represent the \emph{Producer}'s mobility region. To approximate the worst case conditions for the proposed architecture and prevent easy access to \emph{Producer}'s current location, one of the top row ASs--not used by the \emph{Consumer}--is chosen as the \emph{Producer}'s Home Network.}). The selected mobility levels (for the \emph{Random Waypoint Model} in ns3) and their impact on handover frequencies are illustrated in Figure~\ref{figure:handover_results} (single-number scenario refers to constant speed mobility, whereas two-number scenario refers to mobility based on (min,max) speeds). We observe from Figure~\ref{figure:handover_results} that the chosen mobility levels can trigger highly unstable conditions due to frequent handovers, hence, are extremely useful to demonstrate how the three approaches perform under such conditions.

\subsection{Session Throughput}

The CBR model used in ndnSIM assumes an always constant request rate, \emph{i.e.},  no traffic is rushed (including the retransmissions). Hence, throughput performance is inversely proportional to both the recovery latency after handovers (the lower the latency, the higher the throughput) and the retransmitted Interests (the higher the retransmission rate, the lower the throughput). In short, from the perspective of the delay-tolerant traffic, effective throughput represents the percentage ratio of Data packets successfully received by the \emph{Consumer} at the end of a simulation run (\emph{i.e.}, the ratio of the number of Data packets received to the number of Interest packets transmitted by the \emph{Consumer} application).

\begin{figure}[htb]
  \centering
  \includegraphics[width=3.3in]{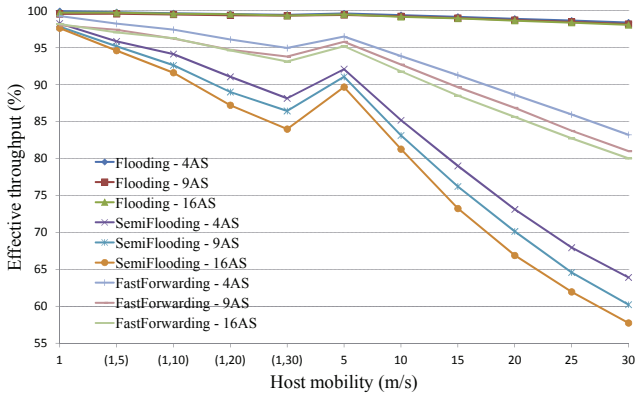}\\
  \caption{Throughput performance of the proposed architecture with respect to \emph{Flooding} and \emph{Semi-flooding}.}\label{figure:results1}
\end{figure}

We illustrate the results for the effective throughput performance in Figure~\ref{figure:results1} (where we use the term \emph{FastForwarding} to refer to our solution). We observe that the proposed framework achieves better than $80\%$ throughput for all the considered scenarios, and performs significantly better than Semi-flooding, regardless of network size. Our analysis suggest that the performance of the proposed architecture can be further improved (to, for instance, better than $90\%$ at the highest user speed for the 9AS scenario) by avoiding retransmissions experienced due to stale PIT entries triggering different paths (that may not reach the \emph{Producer} in time). There are multiple ways to achieve that without introducing significant overhead to the system, for instance, by limiting PIT entries for the FL-based mobile traffic.

\subsection{Recovery Overhead}

\begin{figure}[htb]
  \centering
  \includegraphics[height=3.3in, angle=270]{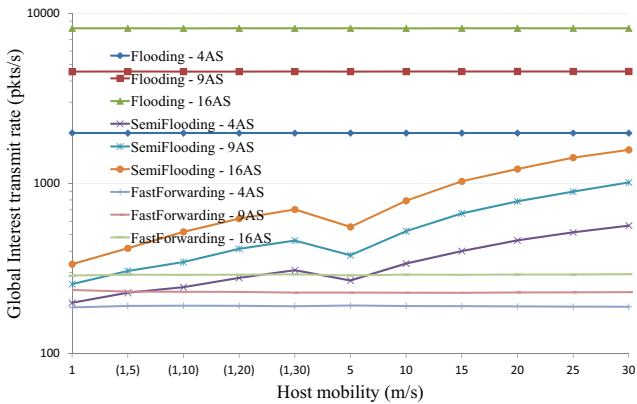}\\
  \caption{Comparative results for the overall network-wide Interest rate.}\label{figure:results2}
\end{figure}

One of our objectives in designing a location-centric forwarding technique is to minimize the overhead introduced during path recovery after mobile handovers. We can clearly see the potential impact of mobile handovers, if we examine the Interest rates shown in Figure~\ref{figure:results2}. We observe significantly better results with our solution that is $5$--$12$ times better, when compared to Flooding, and $2$--$5$ times better, when compared to Semi-flooding.

\begin{figure}[htb]
  \centering
  \includegraphics[height=3.3in,angle=270]{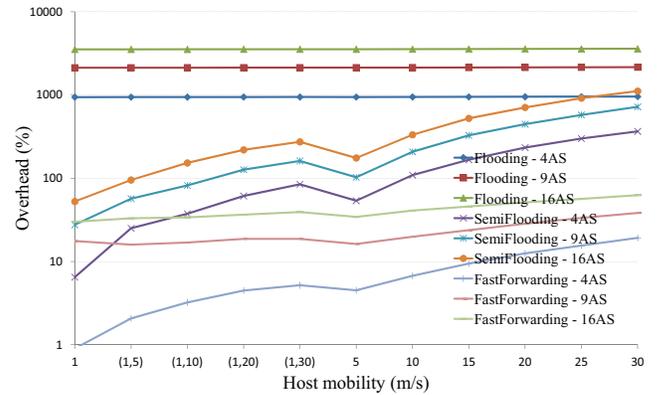}\\
  \caption{Comparative results for the overhead performance (\emph{i.e.}, percentile ratio for the additional Interest packets transmitted to support the perceived Data delivery rates).}\label{figure:results3}
\end{figure}

Another way to look at the overhead performance is to measure the additional number of Interest packet transmissions used to achieve the perceived throughput performance. We show the results in Figure~\ref{figure:results3}. From such perspective, we observe up to $50$--$200$ times better performances when compared to the Flooding technique, and $5$--$20$ times better performances when compared to the Semi-flooding technique, thereby, further illustrating the efficiency of our solution. Also note that, the increase in overhead for our solution is mostly caused by the wasted retransmission attempts leading to lower throughput, as explained earlier. By minimizing the number of such attempts lowers the percentile overhead by 60\%, with much slower-less than half-increase rate between lowest-to-highest mobility levels.

In short, our approach proves to be a much more scalable solution, in both the network size and the host mobility.

Analytically, we can approximate the worst case Interest overhead (corresponding to inter-\textsc{As} handover) for the proposed Mobility Service framework during the lifetime of a session as $O(h\log N)$ (where $h$ represents the overall handover rate, and $N$ represents the network size), which can be determined as follows. During an inter-\textsc{As} handover, the following events take place with our approach: ($i$) \emph{Producer} $P$ registers its prefix to the new remote-domain $D_{C}$, ($ii$) $D_{C}$'s \textsc{Lc} updates $P$'s \textsc{Home-Lc} (\textsc{Hlc}), and ($iii$) $P$'s \textsc{Hlc} updates the previous remote-domain $P$ was registered to, $D_{P}$, which then updates its local \textsc{Er}s and \textsc{Sr}s. The overhead for the first event is given by $O(\delta_{nL})$, for the second event is given by $O(\delta_{nG})$, and for the third event is given by $O(\delta_{nL} + \delta_{nG})$, where $\delta_{nL}$ represents the average distance between two nodes within a local domain, and $\delta_{nG}$ represents the average distance between two nodes within two separate domains. We can approximate the average distance between any two nodes in a network of size $N$ using $\log N$, allowing us to approximate the overhead during a handover using $O(log N)$. On the other hand, for the default NDN forwarding policies (specifically \emph{Smart-flooding} based policies), the overhead during a handover is approximated as $O(N)$, since the handovers trigger \emph{flooding} in the network.

For instance, for the synthetic topologies we considered, going from 4AS to 9AS (or 16AS), $\delta_{nG}$ increases by $1.11$ (or $1.22$). Similarly, going from 9AS to 16AS, $\delta_{nG}$ increases by $1.1$. To represent the change in $N$, we can simply use the ratio of the number of ASs (resulting in $2.25$, $4$, and $1.78$). Our results suggest that, change in overhead for the given scenarios are given by: for \emph{Flooding} ($2.3$, $4.1$, $1.8$); for \emph{Semi-flooding} ($1.4$--$1.8$, $2.2$--$2.8$, $1.3$--$1.6$); and, for our solution, ($1.2$--$1.4$, $1.5$--$1.7$, $1.2$--$1.3$). In short, \emph{flooding}-based techniques observe an increase in overhead closer to change in $N$, whereas our solution observes an increase in overhead closer to $\delta_{nG}$, further proving its scalability.

\section{Discussions}\label{Section:Discussions}
\subsection{Storage Considerations}
FIB represents both the strengths (\emph{i.e.}, flexible operation and on-the-fly routing decisions) and the weaknesses (\emph{i.e.}, overhead) of NDN, with greater perceived impact as the network size and content availability increase. Specifically, ad hoc operation allows for greater flexibility during routing, whereas, increased database size (with variable prefix names and multiple entries per prefix) degrades the operational efficiency at each NDN capable router by increasing the perceived processing latency (see \cite{yuan12} for a detailed discussion on challenges in content-centric forwarding).

FPT addresses these drawbacks by partitioning the information required for end-to-end routing at different sections in the network. The major portion of the -mostly local- FIB entries, \emph{i.e.}, prefix-to-address mappings, are stored at the \textsc{L-db}s. However, due to domain-based partitioning, the number of entries stored at a given \textsc{L-db} is expected to be much smaller than what is expected to be stored in the FIB of an NDN router. \textsc{Sr}s are only responsible for keeping entries of the active hosts they serve, rather than keeping entries for the hosts being serviced by the other routers. \textsc{Er}s provide the backbone dependent mappings and their perceived overhead is limited in the maximum of the number of domains/\textsc{As}s and the number of locally hosted \emph{Producer}s. The intermediate routers, between \textsc{Sr}s and \textsc{Er}s, are only responsible for carrying the next hop-to-\textsc{Er} mappings, hence not anymore observing overhead proportional to the number of hosts being serviced along the downstream channel. As a result, we expect noticeable improvement in the processing latency within the network to route packets between endpoints. Specifically, by forwarding Interests on the \textsc{Lc}-managed address-space, rather than the highly variable prefix-space, lookup latency on a typical NDN router does not anymore depend on the prefix length.
\subsection{Scalability Considerations}
For the proposed architecture, another important concern is the performance of the \textsc{Lc}, which is expected to service the requests of the hosts associated with the \textsc{Lc}'s domain. An \textsc{Lc} carries prefix-to-domain mappings for the hosted \emph{Consumer}s and remote \emph{Producer}s and prefix-to-router mappings for the hosted \emph{Producer}s. Note that prefix-to-domain mappings are updated at a much lower rate (\emph{inter}-\textsc{As} handover rate) than the rate associated with prefix-to-router mappings, which change at the \emph{intra}-\textsc{As} handover rate.
We can therefore approximate the request rate $\rho$ as follows:
\begin{equation}
\rho = h_{AS} \times \big( |mC| + \kappa |RmP|\big) + h_{AP} \times |mP|
\end{equation}
where $\{h_{AS},h_{AP}\}$ represent the mean \emph{inter}-\textsc{As} and \emph{intra}-\textsc{As} handover rate (with $h = h_{AS} + h_{AP}$), $\kappa$ represents the average number of unique \emph{Consumer} domains requesting content from the same \emph{Producer}, $\{mC, mP\}$ represents the sets of mobile \emph{Consumer}s/\emph{Producer}s hosted by the current domain, $\{RmP\}$ represents the set of remotely located mobile \emph{Producer}s using the current domain as a \emph{Home} network, and $|\cdot|$ represents the size operator. As the network size increases, we expect $\kappa$ to converge to a small constant and we can assume $|mC| = |mP| = |RmP| = n_m$. Therefore, we can approximate $\rho$ as $h \times (1 + \gamma \kappa) \times n_m$, where $\gamma = h_{AS}/h$. If we assume frequent handovers with mean inter-handover latency of $10s$, and with $\kappa \approx 1/\gamma$ and $n_M = 1M$, we expect a value of $\rho=200K$ requests per second, which is easily manageable with the current server architectures.

Also note that, because of the overall latency incurred during handovers (caused by handover and end-to-end propagation delays), we consider the requirements on processing latency to be less strict than that of NDN forwarding, thereby allowing the \textsc{Lc} to be more flexible in its lookup operations. Furthermore, if necessary, multiple \textsc{Lc}s can be assigned to a single domain to support load balancing within that domain. In that case, a simple hash function that assigns prefixes to an \textsc{Lc} can be installed on each designated router to support such features. Our framework is flexible enough to manage such scenarios with little increase in complexity.
\subsection{Security Considerations}
There are various ways an attacker can exploit the possible vulnerabilities in our architecture by targeting the \textsc{Lc}s. For instance, by registering non-existing prefixes to the \textsc{Lc} as fake \emph{Producer}s, and by requesting non-existing prefixes from the \textsc{Lc} as fake \emph{Consumer}s, attackers can overload the controllers and limit access to the legitimate requests. We next explain possible approaches we can use in such scenarios to minimize the impact of flooding attacks on the overall performance.
\subsubsection{Producer Flooding}
Our architecture assumes a \emph{Producer} to include certain information within the registration message to identify the \emph{Producer}'s home network and authenticate the registered content. Hence, we can limit the scope of fake-\emph{Producer} attacks through authentication failure messages received from the home networks. After an authentication failure message is received by the host network's \textsc{Lc}, information on the fake-\emph{Producer} can be shared with the host network's \textsc{Sr}s, to prevent or reduce access to the matching user's registration requests.
\subsubsection{Consumer Flooding}
To prevent an attacker from hijacking the network by sending requests for non-existent prefixes, multiple approaches are possible. First, we can employ a threshold-based admission policy at the first point of entry for the incoming requests and limit the number of outstanding requests that await for the path update from the \textsc{Lc}. Our architecture already does this to some extent, by suppressing requests targeting the same entity (\emph{i.e.}, \emph{Producer}) at the \textsc{Sr}s. Second, we can use an adaptive decision policy to enforce stricter threshold values at certain \textsc{Sr}s depending on the experienced overhead at the \textsc{Lc}s. Since the forwarding label in a request message includes information on the entry points, \textsc{Lc}s can aggregate the necessary statistics to quickly determine the problematic areas, and restrict access whenever needed. Third, by sending feedbacks to \textsc{Sr}s on problematic requests, attackers can be identified in a timely manner and the information on them can be shared with other \textsc{Sr}s within the same domain to limit the effectiveness of future attacks.
\section{Conclusion}\label{Section:Conclusion}
In this paper, we proposed a decentralized mobility-centric solution for Named-data Networking (NDN) to address the scalability problems that arise during the delivery of mobile content to requesting \emph{Consumer}s. The proposed solution relies on four essential building blocks to support location-driven forwarding: \emph{Local Controller} (resolution server to provide name-to-locator mappings), \emph{Forwarding Label} (dynamic path information inserted within the Interest to support \emph{iterative-binding} by intelligently guiding the request towards the content source), {\emph{Mobility Tag}s} (single-bit flags inserted within Interest or Data packets to indicate entity mobility or mobility service) and \emph{Fast Path Table} (database utilized at the designated routers to store name-to-locator mappings). We presented an in-depth analysis of the proposed architecture and explained in detail all the necessary steps required to initiate and maintain connectivity between mobile end points. We implemented our solution in ndnSIM and demonstrated significant performance improvements in network scalability while achieving comparable results to \emph{flooding} in effective throughput. We also addressed the practical considerations in regards to storage requirements, controller scalability, and security concerns, and discussed the efficiency and effectiveness of the proposed architecture.

\bibliographystyle{ieeetran}
\bibliography{ICN}

\begin{thebibliography}{10}
\providecommand{\url}[1]{#1}
\csname url@samestyle\endcsname
\providecommand{\newblock}{\relax}
\providecommand{\bibinfo}[2]{#2}
\providecommand{\BIBentrySTDinterwordspacing}{\spaceskip=0pt\relax}
\providecommand{\BIBentryALTinterwordstretchfactor}{4}
\providecommand{\BIBentryALTinterwordspacing}{\spaceskip=\fontdimen2\font plus
\BIBentryALTinterwordstretchfactor\fontdimen3\font minus
  \fontdimen4\font\relax}
\providecommand{\BIBforeignlanguage}[2]{{%
\expandafter\ifx\csname l@#1\endcsname\relax
\typeout{** WARNING: IEEEtran.bst: No hyphenation pattern has been}%
\typeout{** loaded for the language `#1'. Using the pattern for}%
\typeout{** the default language instead.}%
\else
\language=\csname l@#1\endcsname
\fi
#2}}
\providecommand{\BIBdecl}{\relax}
\BIBdecl

\bibitem{SurveyICN}
G.~Xylomenos, C.~Ververidis, V.~Siris, N.~Fotiou, C.~Tsilopoulos, X.~Vasilakos,
  K.~Katsaros, and G.~Polyzos, ``A survey of information-centric networking
  research,'' \emph{IEEE Communications Surveys Tutorials}, vol.~PP, no.~99,
  pp. 1--26, 2013.

\bibitem{ICNsurvey12}
M.~F. Bari, S.~R. Chowdhury, R.~Ahmed, R.~Boutaba, and B.~Mathieu, ``A survey
  of naming and routing in information-centric networks,'' \emph{IEEE
  Communications Magazine}, pp. 44--53, Dec 2012.

\bibitem{MainICNSurvey}
B.~Ahlgren, C.~Dannewitz, C.~Imbrenda, D.~Kutscher, and B.~Ohlman, ``A survey
  of information-centric networking,'' \emph{IEEE Communications Magazine},
  vol.~50, no.~7, pp. 26--36, 2012.

\bibitem{DONA}
T.~Koponen, M.~Chawla, B.-G. Chun, A.~Ermolinskiy, K.~H. Kim, S.~Shenker, and
  I.~Stoica, ``A data-oriented (and beyond) network architecture,'' in
  \emph{ACM SIGCOMM}, 2007, pp. 181--192.

\bibitem{NetInf}
C.~Dannewitz, J.~Golic, B.~Ohlman, and B.~Ahlgren, ``Secure naming for a
  network of information,'' in \emph{IEEE INFOCOM Computer Communications
  Workshops}, 2010, pp. 1--6.

\bibitem{385}
\BIBentryALTinterwordspacing
Z.~Zhu, A.~Afanasyev, and L.~Zhang, ``A new perspective on mobility support,''
  NDN, Technical Report NDN-0013, July 2013. [Online]. Available:
  \url{http://named-data.net/techreports.html}
\BIBentrySTDinterwordspacing

\bibitem{ndnSIM}
A.~Afanasyev, I.~Moiseenko, and L.~Zhang, ``{ndnSIM}: {NDN} simulator for
  {NS-3},'' NDN, Technical Report NDN-0005, Oct 2012.

\bibitem{AzginICC14}
A.~Azgin, R.~Ravindran, and G.~Wang, ``Mobility study for named data networking
  in wireless access networks,'' in \emph{IEEE International Conference on
  Communications (ICC)}, 2014.

\bibitem{AlexThesis}
A.~Afanasyev, ``Addressing operational challenges in {Named Data Networking}
  through {NDNS} distributed database,'' Ph.D. dissertation, UCLA, September
  2013.

\bibitem{yuan12}
H.~Yuan, T.~Song, and P.~Crowley, ``Scalable {NDN} forwarding: Concepts, issues
  and principles,'' in \emph{IEEE International Conference on Computer
  Communications and Networks (ICCCN)}, 2012.

\end{thebibliography}

\end{document}